\documentstyle[preprint,aps]{revtex}
\draft
\tighten
\begin {document}
\title{ Two--Nucleon
       Solitary Wave Exchange Potentials (SWEPs)}
\author {{\Large \bf Mesgun Sebhatu}
\thanks{Presented at a Special Theoretical Physics Symposium to
honor {\bf Philip B. Burt on his 60th Birthday}, Sept. 16, 1994,
University of Georgia, Athens. The author is grateful to George Strobel for
inviting him and to Winthrop Research Council for financial support. }}
\address{Department of Chemistry and Physics
Winthrop University, Rock Hill SC 29733}
\maketitle

\begin {abstract}
The long and intermediate range effects of the strong nuclear force is most
directly representedby a two--nucleon potential.
Six decades of collective research effort has led to the development
of one--boson--exchange--potentials (OBEPs).
 OBEPs are able to account for elastic two--nucleon scattering data.
  However,  they require about a dozen adjustable parameters and rely
on a variety of mesons that are described by linear fields.
 The inclusion of self (meson-meson)
interactions as nonlinearities in the exchanged meson fields has led to a
 new class of two--nucleon potentials  called solitary wave exchange
potentials (SWEPs).
 SWEPs possess few (about three) parameters and yet calculations using
simple cases of SWEPs such as the $\lambda\Phi^4$ and  SG SWEP
 are beginning to yield highquality singlet even two--nucleon
phase shifts.  The long--term goal of this work is to derive
a variety of realistic two--nucleon potentials as generalizations of
the $\lambda\Phi^4$ and sine--Gordon SWEPs that
can play a significant role in the calculation of nuclear, particle, and
astrophysical phenomena.

\end{abstract}

\newpage

\section{ Introduction}
P. B. Burt's work extends to almost all branches of theoretical physics.  In
the last two decades,
I have been fortunate to work with him on the application of  a nonlinear
quantum field
theory he developed \cite{Burt81,Burt74}  to mostly two--nucleon  interactions.
His guidance when I was his student in the early 70s and our collaboration
afterwards has led
to the development of solitary wave exchange two-nucleon potentials (SWEPs). I
am pleased to
present SWEPs on this occasion.

The most popular phenomenological potentials such as the Reid \cite{Reid68}
soft-core and the
best OBEPs such as those developed by Bonn{\cite{Mach87}, Paris \cite{Laco80},
and Nijmegen
\cite{Nag78} groups are all based on the assumption that exchanged mesons
(bosons) are
described by linear fields.
Aside from simplicity, there is no {\em a priori} reason to expect meson fields
to be linear. It
is possible that the exchanged mesons interact with each other and are more
realistically
described by nonlinear fields.
 This possibility has led to the development
of solitary wave exchange potentials(SWEPs) \cite{Seb76,Seb89a},Seb90}
 SWEPs are derived in the same manner as OBEPs \cite{Mach87}. However, OBEPs
rely on a superposition of linear
field propagators that are accompanied with independent masses, coupling
constants and form factors. This approach involves about a dozen
parameters.  In the derivation of SWEPs, the linear field theory based
propagators are replaced with solitary wave propagators constructed from exact
particular
solutions--solitary wave solutions--of nonlinear
generalizations of the Klein--Gordon equation. The solitary wave propagator
exchanges a
sequence of masses with a few (about three) parameters.  The nonlinear quantum
field theory
upon which SWEPs are based is discussed in a monograph by Burt \cite{Burt81}.
SWEPs exhibit
features that are
characteristic of well known phenomenological \cite{Reid68} and one boson
exchange
\cite{Mach87,Laco80,Nag78} potentials. e.g., spin singlet $\ell$ even NN SWEPs
are attractive
at intermediate and long ranges $(r \ge 0.7 fm)$ and repulsive at short ranges
$(r \le 0.7 fm)$.
At long ranges $(r \gg 2 fm)$ all the SWEPs have OPEP tails.

\section{Burt's Nonlinear Quantum Field Theory}
The nonlinear quantum field theory developed by Burt is discussed in his book
\cite{Burt81}. This section barely serves as an intrduction to help establish
the notation used in this paper. It states
the basic nonlinear field equations and the solitary wave solutions relevant to
the SWEPs
discussed in this paper.

The field equations for spin--zero meson fields used in connection with SWEPs
are nonlinear
generalizations of Klein--Gordon equation \cite{Burt81,Burt74,Seb81b}. They
have the form:
\begin{equation}
\partial_{\mu}\partial^{\mu}\Phi + m^{2}\Phi +J(\lambda_{i},\Phi) = 0
\end{equation}
where m is the meson mass (in this proposal the pion mass), $\lambda_i: (i= 1
\ldots.n)$ are
self--interaction coupling constants and J is the meson field self-interaction
current. Simple
examples of eq. 1 are the
$\lambda\Phi^4$ field theory which leads to $J =\lambda\Phi^3$ and the
sine--Gordon equation which follows from $J_sg = m^2/\lambda (\sin
\lambda\Phi) - m^2\Phi$.  These two examples yield essentially equivalent
two--nucleon
potentials \cite{Seb83}.  Even though the major goal of this work is to tackle
SWEPs based on
generalizations of the $\lambda\Phi^4$  and sine--Gordon theory
\cite{Burt81,Seb81b}, in this
paper, the
sine--Gordon equation is used to demonstrate that  realistic
two--nucleon potentials with very few parameters (only one more than OPEP)  can
be derived
from simple nonlinear quantum field theoretical models.
The sine--Gordon equation used is:
\begin{equation}
\partial_{\mu}\partial^{\mu}\Phi + \frac{m^{2}}{\lambda}\sin \lambda\Phi = 0
\end{equation}
In the $\lim_{\lambda \rightarrow 0}$, eq. 2 reduces to the well known
Klein--Gordon equation
\begin{equation}
\partial_{\mu}\partial^{\mu}\phi + m^{2}\phi = 0
\end{equation}
The negative and positive frequency solutions for the Klein--Gordon
equation
(3) are:
\begin{equation}
\phi^{(\pm)}=A_{k}^{(\mp)}\frac{e^{\pm\check{K}\cdot\check{X}}}
                        {(D_{k}\omega_{k})^{-1/2}},
\end{equation}
where $A_{k}^{(\mp)}$ are creation and annihilation operators,$\check{K}$ and
$\check{X}$
are momentum and space four vectors such that
$\check{k}\cdot\check{x} = k_{0}x_{0}-k\cdot{x}$, $\omega_{k}^{2} = k^{2}
+m^{2}$, and
$D_k$ is, in general, a $k$ dependent factor. It plays an
important role when proper normalization is considered \cite{Burt81,Seb89c}. In
this paper,
$D_k = 1$ is used for simplicity. A pair of quantized
solitary wave solutions for the sine--Gordon equation (2) obtained by
direct integration or by the method of base equations
\cite{Burt81} in terms of the linear fields $\phi^{(\pm)}$ is:
\begin{equation}
\Phi^{(\pm)}=\frac{4}{\lambda}\tan^{-1}\left[\frac{\lambda}{4}\phi^{(\pm)}
                             \right].
\end{equation}
Expanding eq. 5 in series
\begin{equation}
\Phi^{(\pm)}=\frac{4}{\lambda}\sum_{n=0}^{N}%
\frac{(-1)^n}{2n+1}\left[\frac{\lambda\phi^{(\pm)}}{4}
              \right ]^{2n+1}.
\end{equation}
Eq. 6 is used to construct the solitary wave propagator \cite{Burt81}
\begin{equation}
 P_{sG}(K^{2};M_{n}^{2})=\sum_{n=0}^{N}[m\beta]^{2n}
       \frac{(2n+1)!(2n+1)^{2n-4}}{K^{2}+M_n^{2}}\Delta_F(K^{2};M_n^{2}),
\end{equation}
where in eq. 7: $M_n = (2n+1)m$; $N$ is the number of terms to be included in
the propagator;
$\beta =\lambda/(16m)$; and $\Delta_F= \{K^{2} -
M_{n}^{2} + i\epsilon\}^{-1}$ is the Feynman meson propagator in momentum
space.
The sine--Gordon solitary wave propagator like other solitary wave propagators
is essentially a
superposition of Feynman propagators and it exchanges a series of meson masses.
The sequence of masses lead to superpositions of attractive as well
as repulsive Yukawa and exponential potentials in coordinate space.

\newpage
\section{Derivation Of Nonstatic SG SWEP} in lowest order, the NN
interaction is conveniently represented by the direct (a) and exchange (b)
Feynman diagrams
shown in Fig. 1 .
At each vertex, following e.g., Buck and Gross \cite{Buck79}, we use a
mixed pseudoscalar (PS) and pseudovector (PV) $\pi$NN coupling of the form:
\begin{equation}
\Gamma = -ig\left[\Omega +\frac{(1-\Omega)k\cdot\gamma}{2M}\right]\gamma_5.
\end{equation}
In eq. (8),  $g$ is the $\pi NN$ coupling constant, $k$ is the exchanged
momentum,  $\Omega$
is the PS-PV mixing parameter.  When $\Omega = 1$,
$\Gamma$ is pure PS and when $\Omega = 0$ it is pure PV. In the static
limit both couplings lead to identical potentials. When leading nonstatic terms
are included,
however, the two types of couplings lead to quadratic spin-orbit terms of
opposite sign.  The
momentum space SG SWEP obtained from the above Feynman diagrams with leading
nonstatic
terms is
\cite{Hosh60,Aris78,Seb81a}:
\begin{eqnarray}
 V(k,q) & = & \frac{g^2}{4\pi} \left[ \frac{m}{2M} \right]^2
           (\tau_1 \cdot \tau_2) P_{sG}(k^2;M^{2}_n) \nonumber\\
     &   & \left \{ \left[1 - \frac{ k^2}{ 4 M^2} \right]
            \left[ \frac{(\sigma_1 \cdot k)(\sigma_2 \cdot k )}{M^2} \right]
\right.  \nonumber \\
     &   & \mbox{} + \frac{2 \Omega - 1}{2 (M m)^2}
            \left [(\sigma_1 \cdot k \times q)
               (\sigma_2 \cdot k \times q) \right.\nonumber \\
     &   & \left. \left. \mbox{} - (\sigma_1 \cdot \sigma_2) (k \times q)^2
\right]
            \right \} .
\end{eqnarray}
Where in eq. 9: $k=p'-p$, $q=(p'+p)/2$, $p(p')$ are the initial (final) momenta
of  the interacting
nucleons; $\tau$ and $\sigma$ are isotopic spin and spin Pauli  spinors. In the
center of
momentum frame, $p_1=p_2=p$ and $p'_1 =p'_2 =p'$.

\section{Nonstatic SG SWEP In Coordinate Space} The coordinate
space SG SWEP obtained by Fourier transforming the momentum space potential (9)
is:
\begin{eqnarray}
 V(X_n )  & = & G(\tau_1 \cdot \tau _{2}) \sqrt{\frac{2}{\pi}}
             \sum_{n=0}^{N} C_{n} X_{n}^n \nonumber \\
       &   &  \left[ (\sigma_1 \cdot \sigma_2 )V_C(X_n)
             + S_{12}V_{T}(X_n)+ (2\Omega-1) L_{12} V_{\ell\ell}(X_n)
             \right],
\end{eqnarray}
where:
\[G = \frac{g^2}{4\pi} \left[ \frac{m}{2M} \right]^2 m; \mbox{ }
   C_n  =\frac{(2n)! \beta^{2n}} {n!2^n };\]
\[V_C (X_n ) = X_{n}^{-1/2}K_{n-5/2}(X_n )-3X_{n}^{-3/2}K_{n-3/2}(X_n );\]
\[V_T
(X_n ) = X_{n}^{-1/2} K_{n-5/2}(X_n );\]
\[V_{\ell\ell}= \frac{1}{2} \left[ \frac{m}{M} \right]^{2}
     \frac{V_T(X_n)}{x^2 };\]
\[L_{12}=(\sigma_{1} \cdot \sigma_2) L^2 -[\sigma_1 \cdot L\: \sigma_2 \cdot L
  +
\sigma_2 \cdot L\: \sigma_1 \cdot L]/2 .\]
In the above equation(s): $X_n =(2n+1)x$, $x=mr$; the terms $V_C$, $ V_T$, and
$V_{\ell\ell}$ are central, tensor, and quadratic spin-orbit  potential
components; and the
$K_{\nu}(z)$ are Bessel functions of the second  kind \cite{Arfk70}.

For spin singlet $NN$ states (total spin $S=0$), the operator $S_{12} = 0$,
and the quadratic
spin operator $L_{12} = -2\ell(\ell+1)$.
The expectation value  of the operator
$ (\tau_1\cdot\tau_2)(\sigma_1\cdot\sigma_2)  = $
$-3$  for singlet even (S=0, T=1,$\ell$ even)  and $+9$ for singlet odd (S=0,
T=0,$\ell$ odd) NN
states.

The SG SWEP (10) can be explicitly written in terms of elementary functions by
using well
known properties of the modified Bessel functions $K_{\nu}(z)$ \footnote{The
information
needed  about $K_\nu (z)$ to simplify eq. 10  when $N\le 4$ is : $K_{\pm
1/2}(z) =\left[ \frac
{\pi}{2z} \right]^{1/2}e^{-z}$;
$K_{\pm 3/2}(z) = (1+\frac{1}{z}) K _{\pm 1/2}(z)$;
and $K_{\pm 5/2}(z)= \frac{3}{z} K_{\pm 3/2}(z)+K_{\pm 1/2}(z)$. }
As an example, the $^{1}S_{0}$ state SG SWEP with the first five (n = 0 to 4)
terms is
shown below:
\begin{eqnarray}
       V_{C}(x) & = &  -G \left[ \frac{e^{-x}}{x} + \beta^{2}(3x-2)
              \frac{e^{-3x}}{3x} \right. \nonumber \\
            &   & \mbox{} + 3\beta^{4}(5x-3)e^{-5x} \nonumber \\
            &   & \mbox{} + 15\beta^{6}(49x^{2}-21x-3)e^{-7x} \nonumber \\
   &
 & \mbox{} + \left. 945\beta^{8}(81x^{3}-18x^{2}-9x-1)e^{-9x}         \right] \
\end{eqnarray}
The first term in (11) is the Yukawa (OPEP) potential and the higher order
$(n>0)$ terms are
modifications due to nonlinear fields introduced by
replacing the Klein--Gordon equation by a sine--Gordon equation.

\section{Some Results} Graphs of singlet even SG SWEPs and phase shifts
obtained from them
shown in Fig. 2  for the leading even $\ell$ spin singlet $NN$
states--$^{1}S_{0}$,
$^{1}D_{2}$ and $^{1}G_{4}$. Fig. 2(a)  shows that the $^{1}S_{0}$ and
$^{1}D_{2}$ state
potentials closely resemble the
corresponding Reid soft-core potentials \cite{Reid68}. Fig. 2(b) shows  the
calculated singlet
even $NN$  SG SWEP phase shifts are in
good agreement with corresponding  experimental values \cite{Arndt}.   In
addition, the SWEPs
yield reasonable $^{3}S_{1} (u)$ and $^{3}D_{1} (w)$ state wave functions (Fig.
3) and
low energy deuteron parameters
[Table I].  These results are remarkable considering only the pion is used and
only two
parameters ($\beta$ and $\Omega$) are involved.  The $^{1}S_{0}$ state is not
affected by the
value of the PS-PV coupling mixing parameter--$\Omega$. Therefore, it involves
only one
parameter $\beta$.  The higher angular momentum states, especially the
$^{1}D_{2}$ state, are
sensitive to the value of $\Omega$.  The values $\Omega= 1/4$, $\beta = 0.89$,
$N =4$,
and $G = 10.35$ MeV, are used in the graphs of the potentials and the
phase shifts shown in Figures 2 and 3 respectively.  The value for $G$
follows from the recent $\pi NN$ coupling constant
 $g^{2}/(4\pi) = 13.55 (f^2 = .075)$ recommended by the Nijmegen Group
\cite{Timm91} , the average  pion
mass $m = 138.033$ MeV and the average nucleon mass $M = 938.926$ MeV  The
elastic
scattering NN phase shifts up to 500 MeV are calculated using the
least--squares method \cite{Seb89b}.
 The least--squares method is more reliable
than conventional methods such as the logarithmic derivative and root search
methods. It has
a built--in means to indicate when the
asymptotic region is reached by providing a measure of the ''goodness of fit''
between the
numerical and asymptotic solution of the radial
Schr\"{o}dinger equation.
It is remarkable that realistic singlet even NN potentials that closely
 resemble Reid soft-core potentials [Fig. 2 (a)] can be derived directly from
nonlinear field theoretical models such as
the sine--Gordon and
$\lambda\Phi^4$ theories. Even though they utilize only one more parameter
($\lambda$ or
$\beta$) than OPEP, the SG SWEP yields high quality phase shifts [Fig. 2(b)] as
well as
reasonable deuteron wave functions [Fig. 3] and low energy parameters [Table
I].  These results
provide a strong
motivation for studying SWEPs and the nonlinear field theoretical models upon
which they are
based.

\section{Conclusion and Future Prospects} The general aim of this work is to
derive  and test
(by comparison with experimental $N N$, $N\overline N$, $\pi N$, and $\pi \pi$
data)  a
variety of two--nucleon potentials called solitary wave exchange potentials
(SWEPs).  The main
virtue of SWEPs is a
strong theoretical foundation that enables them to have less than one third of
the parameters
used by popular OBEPs and yet provide a high quality fit to experimental data.
The immediate
outcome from SWEPs is;
simple, realistic, theoretically derived potentials with the least number of
adjustable
parameters that would yield a full set of high quality
nucleon--nucleon elastic scattering phase shifts as well as low energy
deuteron parameters. The SG SWEP discussed in this paper does not have
 a spin--orbit($L.S$) term. The significance of the $L.S$ term in triplet $NN$
states has
 been long known \cite {Sig58}.
This can be accomplished by incorporating $\rho$ meson, two--pion, and/or
scalar meson exchange
 contribution(s).

Once the parameters of SWEPs are
determined using NN phase shift and low energy deuteron data, They would be
useful in the
quantitative study of nuclear (few and many body),
astrophysical, and hadronic phenomena. The derivation and testing of SWEPs by
comparison
with low and intermediate energy $NN$, $N\overline N$, $\pi N$, and  $\pi \pi$
data has the
important significance of connecting the
nonlinear quantum field theoretical models upon which the SWEPs are based to
experimental
data as well as to QCD inspired quark cluster models \cite {Myhr88}. The
primary aim is,
however, to determine the numerical values of the parameters (self interaction
coupling
constants) involved and establish the reliability as well as usefulness of
nonlinear meson field
theory.  After being tested at low and intermediate energies, the nonlinear
quantum field
theories, especially in their properly normalized form
\cite{Burt81,Seb89c}, can pave the way to dealing with divergence difficulties
that plague
hadronic physics.

\newpage

\begin{figure}
\vspace{3in}
\caption {Second order direct (a) and exchange (b) NN interaction
 Feynman diagrams.}
\end{figure}

\begin{figure}
\vspace{4in}
\caption { (a)  singlet state even $NP$ SG SWEP (solid lines)
        and Reid soft-core potentials (dashed lines); and (b) singlet even  SG
SWEP
$NP$ phase shifts(solid lines) vs Arndt et. al [19]  values (filled circles)}
\end{figure}

\newpage
\begin{figure}
\vspace{3.5in}
\caption{A comparison of $^{3}S_{1} (u)$ and $^{3}D_{1} (w)$ state deuteron
 wave
function using SG SWEP (solid lines) and Reid soft-core (dashed line).}
\end{figure}
\begin{table}
\caption {A comparison of deuteron parameters calculated from
SG SWEP($\beta$ = 0.88) with Bonn OBEP and experimental values [4]. The
normailzation $\int_{0}^{\infty}{(u^2 + w^2)dr} = 0.962$
was used in obtaining the values in the table. The $3.8 \%$  reduction is a
meson exchange current correction.}
\begin{tabular}{lddd}
 Parameter                       & Expt      & SWEP      & Bonn  \\
\tableline
  Quadruple Moment $Q_{d}(fm^2)$  & 0.286     & 0.287     & 0.281 \\
  Magnetic Moment   $\mu_d$       & 0.857     & 0.856     & 0.856 \\
  RMS Radius $r_{d} (fm)$         & 1.966     & 1.9991    &2.002 \\
  D-state probability $P_{D}\%$   & 5$\pm$ 2  & 4.25      & 4.25  \\
\end{tabular}
\end{table}

Figures will be sent by mail upon request.\\
Send requests for figures to:\\
\vspace{6in}

sebhatum@lurch.winthrop.edu.

Please include your mailing address in the E-mail.

\end{document}